# Compositional Diversity Among Primitive Asteroids

Humberto Campins[1], Julia de León[2], Javier Licandro[2], Amanda Hendrix[3], Juan A. Sánchez[3] and Victor Ali-Lagoa[4]

[1] University of Central Florida, Orlando, Florida, USA.
campins@physics.ucf.edu
[2] Institute of Astrophysics of the Canaries, Tenerife, Spain
[3] Planetary Science Institute, Tucson, Arizona, USA
[4] Max Planck Institute for Extraterrestrial Physics, Garching, Germany

**ABSTRACT.** Spectroscopic observations from the ultraviolet to the mid-infrared have revealed new and diagnostic differences among primitive asteroids. We review the spectral characteristics of these asteroids and their inferred compositional and physical properties. Primitive asteroids throughout the belt show carbon-rich compounds, varying degrees of aqueous alteration and even surface ice; recent observations provide significant new constraints on composition, thermal inertia, and other surface properties. New mid-infrared connections between primitive asteroids and interplanetary dust particles indicate that the latter sample a larger fraction of main belt asteroids than meteorites. Links with the composition of comets are consistent with a proposed continuum between primitive asteroids and comets. Two sample-return missions, OSIRIS-REx and Hayabusa 2, will visit primitive near-Earth asteroids (NEAs). Most spacecraft-accessible NEAs originate in the inner asteroid belt, which contains several primitive asteroid families and a background of primitive asteroids outside these families. Initial results from these families offer a tantalizing preview of the properties expected in the NEAs they produce. So far, primitive asteroids in the inner belt fall into two spectral groups. The first group includes the Polana-Eulalia families, which show considerable spectral homogeneity in spite of their dynamical and collisional complexity. In contrast, the Erigone and Sulamitis families are spectrally diverse and most of their members show clear 0.7-μm hydration features. The two sample-return targets (101955) Bennu and (162173) Ryugu, most likely originated in the Polana family.

**Keywords:** primitive asteroids, near-Earth asteroids, asteroid families, meteorites, carbonaceous chondrites, spectroscopy.

## 5.1 Introduction

Primitive asteroids are of interest for a number of scientific and practical reasons. These objects are the likely sources of the least-altered meteorites, the carbonaceous chondrites, and hence, contain clues to the original composition and evolution of our Solar System. Primitive asteroids also represent hazards and resources to humans. Many primitive asteroids have been classified as "potentially hazardous", which means that they are not currently a threat to Earth, but they could become one as their orbits evolve. Several of these potentially hazardous asteroids are the targets of space missions, including two current sample return missions: OSIRIS-REx and Hayabusa 2. Because of the composition and accessibility of primitive asteroids, there is a growing interest in them for in-situ resource utilization (ISRU). The cost of extracting fuel and building materials from asteroids may soon be lower than the cost of launching these resources from Earth, thus creating the foundation for a self-sustaining space economy. Among the goals of both sample-return missions is the identification of resources for future exploitation of asteroids.

In this chapter, we define primitive asteroids as those with low reflectivity (i.e., visible geometric albedo $\leq$ 0.15) and mostly featureless spectra in the visible. In the Tholen classification system, which considers albedo and spectra, primitive asteroids are those in the C-complex, which include the B, C, D, F, G, and T types (Tholen and Barucci 1989). Our understanding of the composition of primitive asteroids had been limited by the paucity of diagnostic spectral features at visible and near-infrared (VNIR) wavelength, where most spectroscopic observations have been made (e.g., Ziffer et al. 2011; de León et al. 2012). However, recent observations at wavelengths ranging from the ultraviolet to the mid-infrared (0.2 to 40 μm) reveal new and diagnostic spectral differences among primitive asteroids (e.g., Rivkin et al. 2011; Hendrix et al., 2016b; Fornasier et al. 2014; Landsman et al. 2016). In this work, we review the spectral characteristics of primitive asteroids and their inferred surface compositions. More detailed discussions of asteroid compositions and their links to meteorites can be found in the chapters by Cloutis et al., Nuth et al., Takir et al. and Zolensky et al. in this volume.

## 5.2 Primitive Asteroid Locations

Primitive asteroids are present throughout the asteroid belt. They are most abundant in the outer belt and Jupiter-Trojan clouds, but are also present in the inner belt and the near-Earth asteroid (NEA) populations with abundances roughly equal to those of higher albedo asteroids. Primitive inner belt asteroids are of special interest as they appear to be the main sources of primitive NEAs and of carbonaceous meteorites (e.g., Bottke at al. 2002; Campins et al. 2010, 2013). A dramatic illustration of how NEAs can enter our atmosphere and become meteorites was the discovery of 2008 TC$_3$ prior to entering Earth's atmosphere and producing the Almahata Sitta meteorites (e.g., Jenniskens et al. 2009; Binzel et al. 2015). Interestingly, it is estimated that 99.9% of the mass of 2008 TC$_3$ was lost during atmospheric entry. So, it is conceivable that the study of primitive asteroids will reveal components that do not survive atmospheric entry, and are therefore not found in primitive meteorites. This is one reason for the continued study of primitive asteroids using Earth and space based techniques.

Primitive NEAs have and continue to be targeted by spacecraft, including robotic and human space missions (e.g., Abell et al. 2015, Mazanek et al., 2016). The two primitive NEAs that are about to be visited by sample-return missions are (101955) Bennu, target of NASA's OSIRIS-REx (Lauretta et al., 2010), and (162173) Ryugu, target of JAXA's Hayabusa2 (Tsuda et al., 2014). Both missions will rendezvous their respective asteroids in 2018. In addition, NEA (65803) Didymos is the target of the NASA's Double Asteroid Redirection Test (DART) mission; this asteroid is classified as Xk, and is not yet clear if it is primitive. Another primitive NEA of special interest is 2008 EV5 because its orbit is particularly accessible from Earth; hence this object has been considered as a spacecraft target by two missions and will likely be visited by a spacecraft in the future (e.g., Mazanek et al., 2016). All four of these NEAs have been classified as potentially hazardous asteroids.

### 5.2.1 Sources of NEAs

The orbits of NEAs have short lifetimes ($\leq 10^7$ years) compared with the age of the Solar System, so these objects originated elsewhere and were transported to their current orbits by a replenishment mechanism. Mechanisms that produce asteroid and planet migrations are also relevant to understanding the mixing of material throughout the history of our Solar System. Most NEAs come from the main asteroid belt and almost all taxonomic classes observed in the main-belt are also found in the NEAs population (e.g., Binzel et al. 2015). Dynamical models and spectral comparisons show that all the spacecraft-targeted NEAs were likely delivered from low-inclination (i < 12°), inner-asteroid belt (2.15 < a < 2.5 AU) populations, including asteroid families and the background outside those families (e.g., Campins et al., 2010, 2013; Bottke et al., 2015). Understanding the main belt sources of primitive NEAs targeted by spacecraft improves our ability to predict their surface characteristics and provides a framework for interpreting the *in situ* and sample-return results, thus integrating smaller-scale details with the bigger picture of asteroid and Solar System origin and evolution.

The dynamical method described in Bottke et al. (2002) continues to be an effective tool for estimating the origin of current NEA orbits; this model is being refined (Bottke et al. 2014), but the results are in good agreement with Bottke et al. (2002). In this model, the authors numerically integrated the orbits of thousands of test particles, starting from the five most efficient source regions of NEAs. These source regions for test particles are: (a) the $\nu_6$ secular resonance at ~ 2.15 AU, which marks the inner border of the main belt; (b) the Mars-crossing asteroid population, adjacent to the main belt; (c) the 3:1 mean-motion resonance with Jupiter at 2.5 AU; (d) the outer main-belt population between 2.8 and 3.5 AU; and (e) the Jupiter-family comets. According to the Bottke et al. (2002) model, about 60% of NEAs come from the inner-belt (a< 2.5 AU), so it is quite possible that all four spacecraft-targeted NEAs (described above) are likely from the inner-belt. For this reason, we give special attention to inner-belt asteroids in this section; however, in later sections we also discuss primitive asteroids elsewhere in the belt, including those in the Pallas, Themis and Veritas families.

### 5.2.2. Inner Belt Primitive Asteroids: Families and Background

Within the inner belt, the sources of NEAs and meteorites are asteroid families and a background of asteroids outside those families. Most NEAs are small asteroids (diameters ≤ 1km), which are unlikely to be primordial because their collisional lifetime is much shorter than the age of our solar system (e.g., Bottke et al. 2005); thus, they are believed to be the fragments of larger objects. Asteroid families and any collisionally-evolved population of asteroids will yield small fragments in this size range; this happens either during the family-forming event (a disruption of a large asteroid tens to hundreds of kilometers in size) or during the collisional evolution that normally occurs in the asteroid belt.

The origin and evolution of inner belt families is an active area of study (e.g., Dykhuis and Greenberg, 2015; Bottke at al. 2015; Reddy et al. 2014; Campins et al. 2013, and references therein). As of late 2017, seven primitive asteroid families have been identified in the low-inclination region of the inner belt (e.g., Nesvorný et al., 2015). These are the Chaldaea, Clarissa, Erigone, Klio, Polana (including Eulalia, which is within Polana) and Sulamitis families (Figure 1, left panel). In addition to these low-albedo families identified so far, there is a population of low-albedo asteroids left after the family members have been subtracted (Figure 5.1, right panel). This background population extends to the $\nu_6$ resonance and it may be as important a source of NEAs as the families (e.g., Campins et al. 2013). In Figure 5.1 the families clearly extend beyond the specific dynamical boundaries chosen by Nesvorný et al. (2015), and it appears that these families are an important source of the background. However, there may also be a primordial component to the background (Delbo et al. 2017). There is only one intermediate-albedo family in the inner belt, Baptistina. This family has a mean geometric albedo of 0.16 (Nesvorný et al. 2015) and is located near the $\nu_6$ resonance. The spectra of the Baptistina members indicate they are not primitive (Reddy et al. 2014). Hence, we focus on the low-albedo families and background asteroids in the inner-belt.

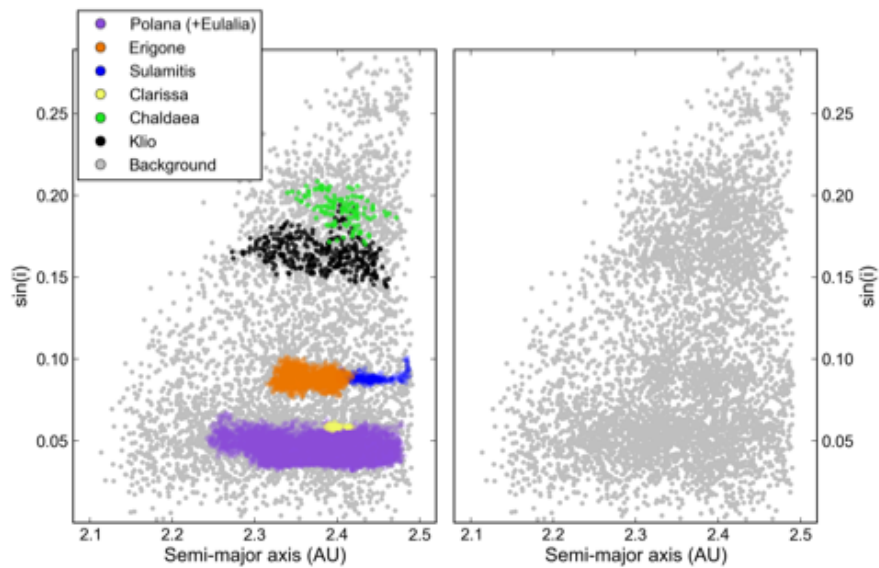

**Figure 5.1**. Adapted from Campins et al. (2013). Left: the distribution of low-inclination, low-albedo inner-belt families and the background population in the space of proper semi-major axis and inclination. Right: The "background" population of low-inclination, low-albedo asteroids left after the family members have been subtracted. Clearly the families extend beyond the specific dynamical boundaries chosen by Nesvorný et al. (2015), and it appears that the families are an important source of the "background".

Several published studies have characterized the spectral features of the largest primitive, low-albedo, inner-belt families, the Polana-Eulalia and the Erigone families (Campins et al. 2010; Campins et al. 2013; Walsh et al. 2013; Bottke et al. 2015; de León et al. 2016; Morate et al. 2016; Pinilla-Alonso 2016). There are at least five inner-belt populations of primitive asteroids (four families and the background) that have yet to be studied in detail (e.g., Morate et al. 2018).

These publications indicate that inner-belt primitive families fall into two spectral groups: Erigone-like (hydrated and spectrally diverse) and Polana-like (no 0.7-μm hydration feature, see Section 5.4, and spectrally homogeneous). More specifically, there is spectral homogeneity within the Polana-Eulalia family complex and sharp differences with the Erigone family (Figures 5.2 and 5.3). The dynamical and collisional complexity of the Polana-Eulalia family (e.g., Bottke 2015; Dykhuis and Greenberg 2015; Walsh et al. 2013) does not translate into the spectral diversity that Walsh et al. (2013) predicted. In contrast, Morate et al. (2016) found that the distribution of taxonomic classes within the Erigone family (Figure 5.3) differs significantly from that of Polana-Eulalia. Morate et al. (2016) showed that the majority of the Erigone asteroids show a 0.7-μm hydration feature, while the Polana-Eulalias do not. These differences have implications for whether or not hydrated minerals may be present on the surfaces of Bennu, Ryugu, and 2008 $EV_5$, all three of which could come from these two families. In fact, a combination of the dynamical and spectral evidence, favors the Polana-Eulalia complex as the most likely origin of Bennu and Ryugu (de León et al. 2018). It has been suggested that the paucity of hydration features in the Polanas may be related to the family's greater age and it could be due to space weathering (Section 5.6). If so, hydrated minerals could still be present just below the surface, even when no 0.7-μm hydration feature is detected (Section 5.4).

In summary, primitive asteroids in the inner-belt and in near-Earth space are especially relevant to predicting the surface characteristics of current and future spacecraft targets. The study of these asteroids can help with spacecraft and mission design, mission operations, and sample site selection criteria. In addition, the spectroscopic characterization of asteroids is already helping to identify targets with the desired composition for ISRU, such as those rich in hydrated minerals, those that are metal-rich, and even some that are both (e.g., Landsman et al. 2015).

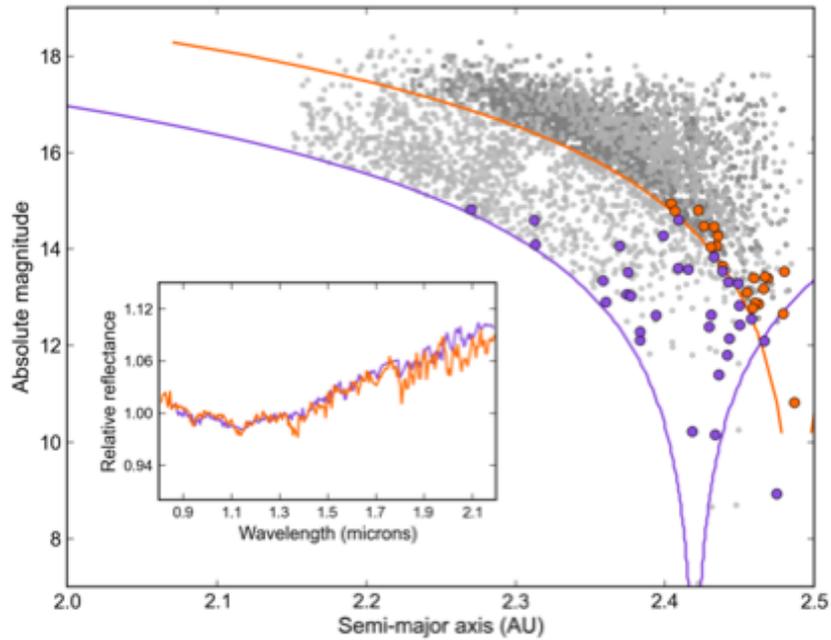

**Figure 5.2**. Spectral homogeneity within the Polana-Eulalia complex, adapted from two figures in Pinilla-Alonso et al. (2016). The outer frame shows the absolute magnitude "H" of the Polana and Eulalia families as a function of the proper semimajor axis "a". The violet and the orange lines are the Yarkovsky cones proposed by Walsh et al. (2013) for the Polana (light grey dots) and Eulalia families (dark grey dots), respectively. The inner frame shows that the two mean spectra of Polana and Eulalia family asteroids are indistinguishable, suggesting a clear compositional homogeneity in the region. The spectra have been normalized to unity at 1μm.

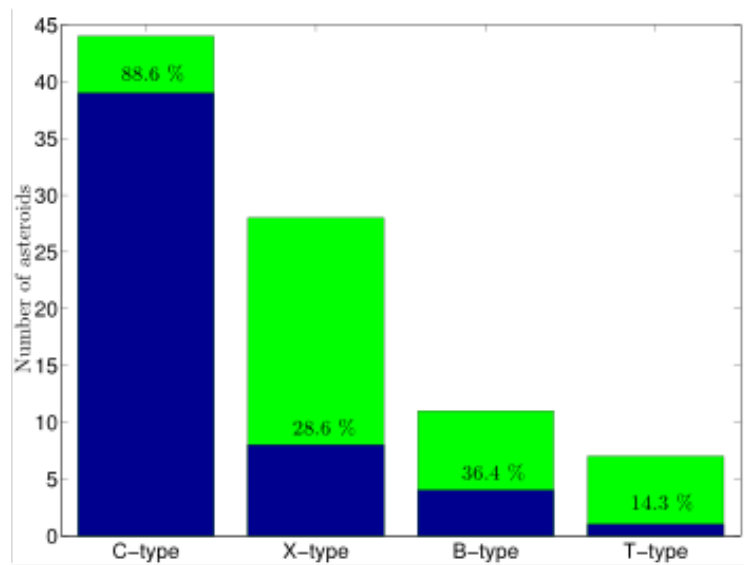

**Figure 5.3** From Morate et al. (2016). The different spectral types within the Erigone family, and the percentage of asteroids showing the 0.7-μm hydrated silicate absorption band (blue), for each primitive taxonomic class (green).

## 5.3 Ultraviolet Spectra of Primitive Asteroids

In the ultraviolet (henceforth UV), primitive asteroids have been observed using the Hubble Space Telescope (HST), the International Ultraviolet Explorer (IUE) and other space observatories, and these measurements have been supplemented with a modest set of lab experiments. At UV wavelengths, ranging from 0.1 to 0.4 µm, the strong constraints imposed by the Earth's atmosphere (mainly the UV absorption by ozone) have generally prohibited large-sample studies of asteroids in this wavelength region. However, the existing observations have shown that primitive asteroids are rich in diagnostic spectral features in the UV and that we are far from fully understanding the information they provide. From 1978 to 1992, the IUE targeted 45 asteroids (including 23 primitive asteroids), generating what remains to date the largest published sample of UV (0.23-0.32 µm) asteroid spectra (Roettger and Buratti 1994). The main finding of that study was that the two major taxonomic classes defined in the visible wavelength range (S-complex and C-complex) persist into the UV. A more recent work by Waszczak et al. (2015) presented photometry of 405 asteroids observed serendipitously by the GALEX space telescope from 2003 to 2012 using its UV filter (0.18-0.28 µm). Their derived UV-V color distribution confirms the results obtained from the IUE data: S-types are redder in the UV than C-types.

The dark appearance (low albedo) of primitive asteroids has been attributed to comparatively high abundances of carbonaceous compounds. Direct identification of such compounds has been challenging because existing studies of organics used visible and near-infrared (VNIR) data (Cloutis et al. 1994). However, carbon-rich species are spectrally active in the ultraviolet, from the far-UV region (FUV, ~0.1-0.2 µm) to the near-UV (NUV, ~0.2-0.4 µm). In fact, laboratory measurements of terrestrial minerals and meteorite samples have shown many potentially diagnostic absorption features in the UV (Cloutis et al., 2008; Hendrix et al., 2016a). The shape of the diagnostic features produced by carbons in the UV depends on several factors, such as the size of the particles, the way they are clustered, and the amount of processing they have undergone, i.e, ion bombardment, solar wind irradiation, heating, etc (e.g., Schnaiter et al. 1998). One of the most prominent features is the one identified as an extinction "bump" in the UV spectra of interstellar medium (Fig. 5.4), centered at ~0.2 µm (Stecher 1965), which has been tentatively associated with graphite grains (e.g., Ferriere 2001), hydrogenated amorphous carbon, or even silicate particles covered by polycyclic aromatic hydrocarbons or PHAs (Henning and Schnaiter 1998). This 0.2 µm feature narrows and shifts to longer wavelengths as carbon species are more and more processed (Papoular et al. 1996); another interesting spectral behavior is the rise in the FUV from 0.2 to 0.1 µm (Figure 5.4). Hendrix et al. (2016a) suggested that a potential technique for studying exposure age of carbon-rich bodies in the solar system is to observe their UV spectrum to characterize the development of the 0.2 µm feature and FUV rise.

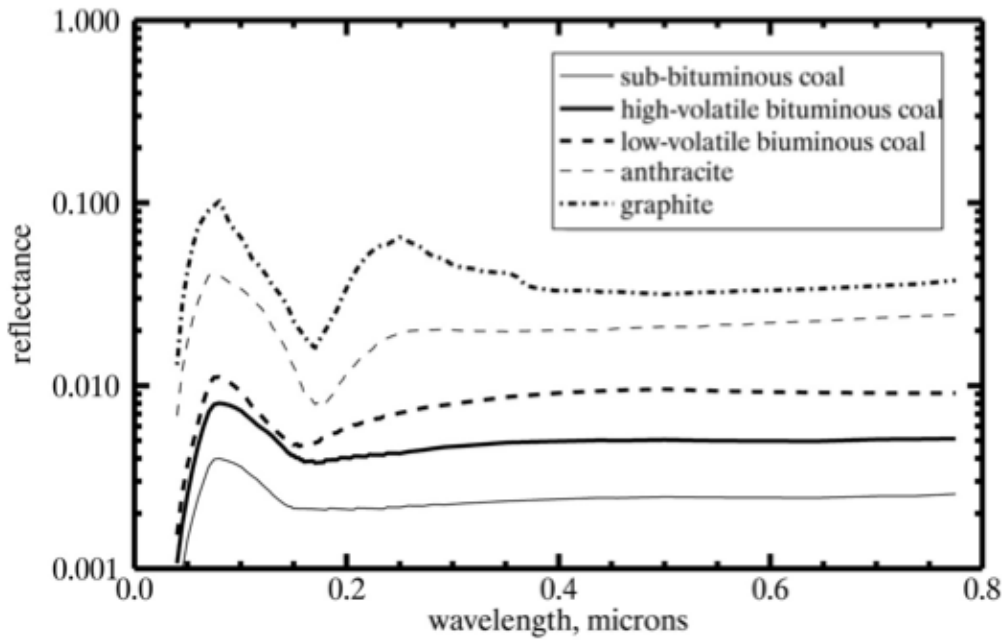

**Figure 5.4**. Coals with increasing graphitization, showing that the absorption feature near 0.2 μm becomes stronger and narrower and shifted to the red; after Papoular et al. (1995) (their fig. 1); spectra are offset. The rise in the FUV from 0.2 to 0.1 μm is most pronounced in the graphite. From Hendrix et al. (2016a)

The few UV spectra of asteroids obtained so far show remarkable diversity, including the existence of an absorption band (centered at ~ 0.28 μm) in the case of Ceres (Rivkin et al. 2011) with an apparent strong FUV upturn in reflectance (Hendrix et al., 2016b), and the absence of one in the spectrum of asteroid Šteins (A'Hearn et al. 2010). Substantial spectral differences among different asteroids in the UV need to be confirmed. Hendrix et al. (2016a) proposed that a "carbon-continuum" exists throughout the Solar System, where the more evolved (i.e. heavily processed) carbons in the inner Solar System exhibit a stronger UV absorption feature and associated FUV rise.

Spectral features indicative of hydration are evident in many primitive asteroids, and this is an area in which the UV spectral range can potentially help. In many asteroids, the UV dropoff (i.e., the UV absorption edge near 0.4-0.5 μm) has long been attributed to a ferric oxide intervalence charge transfer transition (IVCT; e.g., Vilas and Sykes, 1996), which is common in iron-bearing silicates (Gaffey, 1976). In Ceres and in carbonaceous chondrites the dropoff has been identified with phyllosilicates (e.g., montmorillonite; Johnson and Fanale, 1973). The phyllosilicate identification is most closely associated with the 3 μm spectral feature indicative of adsorbed and interlayer water (e.g., Lebofsky, 1978; Section 5.4). The strength of the UV dropoff has been related to heating up to ~700°C (Vilas and Sykes, 1996; Hiroi et al., 1996), connecting water of hydration in phyllosilicates with location and strength of the UV absorption (e.g., Gaffey and McCord, 1978). Furthermore, a 0.7 µm feature is due to hydrated species, and is seen in ~50% of C class asteroids (Section 5.4). It is believed that the presence of opaque materials in the surfaces of low-albedo, primitive asteroids can mask the IVCT feature in the 0.50-0.75 µm region and slightly lowers the absorption in the NUV-visible region (Fornasier et al. 2014). This is something that remains to be tested as more primitive asteroids are observed in the UV.

Finally, a notable feature at 0.43 µm was first discussed by Vilas et al. (1993; Fig. 5.5). Ferric iron and the resultant band at 0.43 µm, is created as a result of aqueous alteration, and is often seen in conjunction with absorption features attributed to phyllosilicates as a result of aqueous alteration activity (Vilas et al., 1993). Fornasier et al. (2014) also detect this absorption band in some of the spectra of their sample of observed asteroids. However, the fact that the solar spectrum presents an absorption band centered at 0.43 µm, and that many of the solar analogue stars commonly used to obtain the reflectance spectra of asteroids have a spectral behavior different from that of the Sun at wavelengths shorter that 0.5 µm, poses some doubts on the interpretation of the 0.43-µm absorption feature and demands a more exhaustive study.

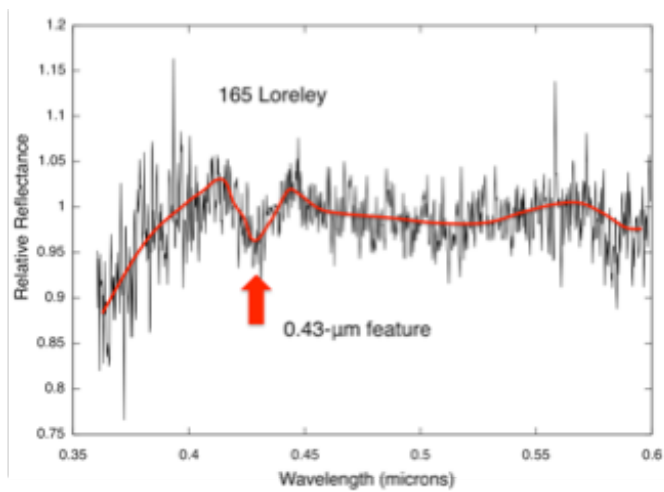

**Figure 5.5**. Near-ultraviolet and visible spectrum of asteroid (165) Loreley from Vilas et al. (1993). The feature associated with Ferric iron and centered at 0.43-µm has been highlighted in red.

### 5.4 Visible and Near-Infrared Spectra

As mentioned in previous sections, the reflectance spectra of primitive asteroids show relatively few features, all of which appear to be associated with the presence of hydrated minerals. In addition to the ultraviolet spectra discussed above, four other observing windows are commonly used to characterize the spectra of asteroids: the visible (~0.5 – 0.9 µm), the near-infrared (0.8 – 2.5 µm), the so-called 3-µm region (2-4 µm), and the mid-infrared (5 – 40 µm). In this section we review the visible and the near-infrared wavelength ranges, as well as the 3-µm region.

### 5.4.1 The visible region

The most prominent feature in the visible spectra of primitive asteroids is the 0.7-µm feature (Fig. 5.6), attributed to $Fe^{2+}$-$Fe^{3+}$ intervalence charge transfer and associated to phyllosilicates (Vilas and Gaffey 1989). Meteorites that are thought to originate from primitive asteroids contain abundant phyllosilicates. The wavelength position of the center of this absorption band might vary from 0.59-0.67 µm for the saponite mineral group to 0.70-0.75 µm for the mixed valence Fe-bearing serpentine group (Cloutis et al. 2011). A more detailed description of the spectral properties of hydrated silicates is provided in the chapter by Cloutis et al. in this volume.

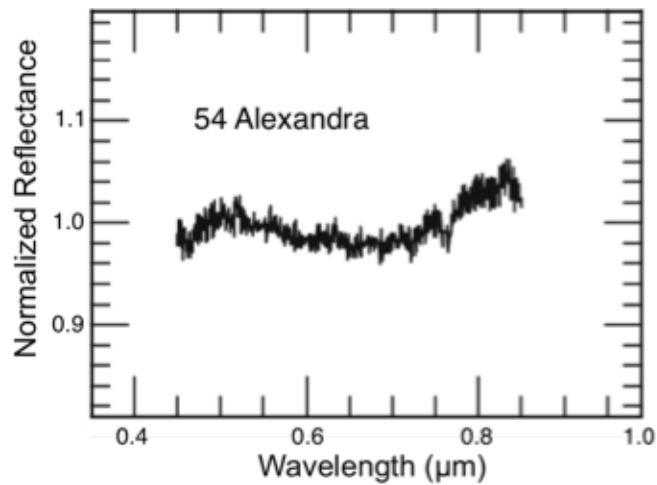

**Figure 5.6**. Adapted from Fornasier et al. (2014). Visible spectrum of asteroid (54) Alexandra, showing the 0.7-µm absorption feature associated to phyllosilicates.

One of the most exhaustive studies regarding the 0.7-µm feature is the one from Fornasier et al. (2014). They analyzed the visible spectra of a total of 600 asteroids classified as C, G, F, B and P following the Tholen taxonomy, finding that about 50% of the C-types and all the G-type asteroids revealed features suggesting the presence of hydrous materials. They concluded that the aqueous alteration sequence starts from the P-type objects, practically unaltered, and increases through the P – F – B – C – G asteroids, these last being widely altered. Their results are in good agreement with those from Vilas (1994), who found that 48% of the C-types showed the 0.7-µm absorption band. Both Fornasier et al. (2014) and Vilas (1994) found that aqueous alteration processes dominate in primitive asteroids located between 2.1-2.6 and 3.1-3.5 AU and that the proportion of asteroids with this band increases with increasing diameter. In particular, Fornasier et al. (2014) showed that the aqueous alteration process is less evident for bodies smaller than 50 km, while it dominates in the 50-240 km sized primitive asteroids. This last point regarding the size is also observed by Carvano et al. (2003) and further discussed there: similarly to what was argued by Vilas and Sykes (1996), the apparent lack of asteroids with the 0.7-μm band at smaller diameters could be explained by a scenario where the larger asteroids would be the remnants of larger parent bodies that underwent thermal and aqueous alteration, while the smaller asteroids would not have been aqueously processed.

In stark contrast with these observations of aqueous alteration primarily in large asteroids is the case of the Erigone and Sulamitis primitive collisional families, which are located in the inner belt. Morate et al. (2016; 2018) studied these two families with the 10.4m Gran Telescopio Canarias, which is powerful enough to sample main-belt asteroids with diameters smaller than 10km. For the Erigone family these authors presented visible spectra of 101 asteroids. They found that C-, B-, X, and T-types within this family show the 0.7-μm absorption band in different percentages: 89%, 36%, 29%, and 14%, respectively (Figure 5.3); similar results were observed in the Sulamitis family (Morate et al. 2018). With the exception of the largest asteroid in the Erigone family, (163) Erigone, which also shows the hydration feature, all of the observed members have diameters smaller than 10 km. Clearly, aqueous alteration is not limited to large asteroids, at least not in the Erigone and Sulamitis families. It is not apparent why smaller asteroids in other populations show a paucity of 0.7-μm absorptions (e.g., Fornasier et al. 2014). Interestingly, the 0.7-μm absorption is absent in all members of the nearby Polana-Eulalia primitive family, which is significantly larger and older than the Erigone family; this family was studied with the same techniques by the same team as the Erigone and Sulamitis families (de León et al. 2016). Hence, it appears that size is a factor in the hydration of some primitive asteroids (primarily in the outer belt), but not in others. Additional factors may include the initial composition and size of the precursor asteroid, space weathering, etc.

A dramatic contribution to asteroid visible spectroscopy will come from the European Space Agency's Gaia mission. Launched in December 2013, Gaia will obtain low-resolution spectra of approximately 300,000 asteroids, covering the wavelength range from 0.35 to 0.9 μm, and therefore sensitive to the 0.7-μm feature. The Gaia asteroid spectral data will be released in 2019 and will result in a major increase in the number and quality of asteroid spectra. This homogeneous data set, with high sensitivity in the near-ultraviolet, will complement nicely ground-based studies and may provide new and diagnostic links between main belt asteroids and NEAs. This may be especially so for C and B-type asteroids, which have indicative absorptions in the near-ultraviolet (e.g., de León et al. 2012). Significant contributions are also expected in the understanding of asteroids families; for example, spectra of main belt asteroids as small as 2 km in diameter will constrain or determine the composition and thermal evolution of a family's parent body. In the case of the Themis family, one of the largest and oldest primitive families, Gaia's spectra will also help establish the role of water and hydrated minerals in the puzzling nature of the "activated asteroids" (also known as main belt comets) within this family (Fornasier et al. 2016; Campins et al. 2012).

### 5.4.2 The near-infrared region

Although the near-infrared spectra of primitive asteroids lack the prominent absorption features common in S-complex asteroids, some spectral variability is indeed found, with slopes ranging from blue to very red and opposite concavities (e.g., de León et al. 2012; Ziffer et al 2011). A dramatic illustration of this variation is the case of the B-type asteroids, presenting a negative (blue) spectral slope in the visible wavelength range. Asteroids with blue spectral slopes are not very common, and it is unclear what minerals or mechanisms produce such blue slope. In 2012, de León et al. presented a study analyzing the visible and near-infrared (0.4 – 2.5 µm) spectra of a total of 45 B-type asteroids, many of them members of the Pallas and the Themis collisional families. A previous study by Clark et al. (2010) with a smaller sample of 22 B-type asteroids had suggested that they fall into three groups: (i) B asteroids with negative spectral shapes, like asteroid (2) Pallas; (ii) B asteroids with concave up curve shapes like asteroid 24 Themis; and (iii) everything else. On the other hand, the results from de León et al. (2012) showed that, instead of the three groups defined by Clark et al. (2010), there was a continuous shape variation in the 0.8 – 2.5 µm range, from a monotonic negative (blue) slope to a positive (red) slope (Fig. 5.7). Most of the Themis family spectra in Clark et al. (2010) had been obtained by Ziffer et al. (2011) and were used to do a comparison with the near-infrared spectra of another primitive family, Veritas, that showed the opposite behavior with a concave down shape and a negative slope in the 1.6 – 2.4 µm region.

Many of the B-type asteroids in de León et al. (2012) had VNIR spectra compatible to carbonaceous chondrites showing different degrees of aqueous alteration. Their conclusions were confirmed by Alí-Lagoa et al. (2013) using 3.4-µm data from the Wide-field Infrared Survey Explorer (WISE). More detailed comparisons between asteroids and meteorites are discussed in the chapter by Cloutis et al. in this volume.

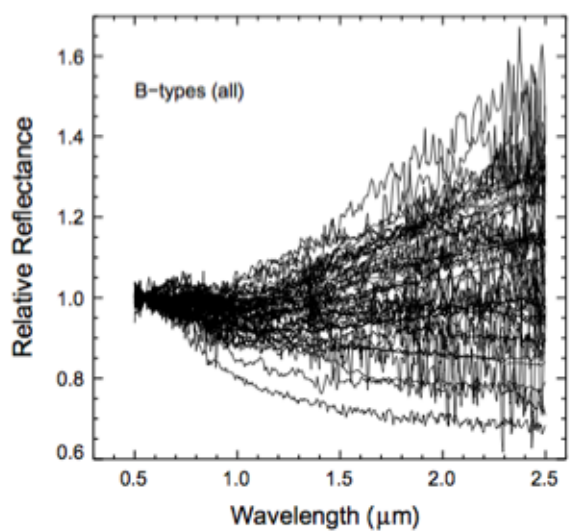

**Figure 5.7**. Visible and near-infrared spectra of a total of 45 B-type asteroids. The spectra are uniformly distributed filling the gap between blue and red slopes in the near-infrared. Adapted from de León et al. (2012).

### 5.4.3 The 3-μm region

In the case of the 3-μm region, water ice, water-bearing materials, and the surficial OH (hydroxyl) from the solar wind have strong absorptions as well. Different and complementary studies have established that the spectra of carbonaceous chondrites present a strong absorption edge near 2.7 μm: the band minimum in the 2.7-2.8-μm region is indicative of phyllosilicate composition and degree of aqueous alteration (Takir et al. 2013). Although the 3-μm is by far the most diagnostic it is also among the most challenging regions in terms of observational constraints, as the thermal background is high and so is the absorption due to water vapor in Earth's atmosphere; typically ground-based data in the 2.5-2.8 μm region are omitted. Therefore, the visible wavelength region and the presence or absence of the 0.7-μm absorption feature is used as a valuable proxy for hydration: different studies show that whenever the 0.7-μm band is present, the 3-μm band is also seen; the inverse is true in only half of the cases (e.g., Vilas 1994; Rivkin et al. 2002, 2015). Hence, a lack of a 0.7-μm feature does not necessarily mean that a surface is anhydrous. A more detailed discussion of spectroscopy in the 3-μm region is found in the chapter by Takir et al. in this volume.

We expect very significant contributions to asteroid spectroscopy from NASA's James Webb Space Telescope (JWST). In this wavelengths range, JWST's Mid-Infrared Instrument (MIRI) will provide low-resolution ($\lambda/\Delta\lambda \sim 100\text{-}300$) and medium-resolution (~1000) spectroscopy from 0.7 to 5.3 microns.

### 5.5 Mid-Infrared Spectra

The spectral information in the mid-infrared region (MIR) of asteroids confirms and complements studies at other wavelengths. The emissivity spectra of asteroids in the MIR are well suited to addressing silicate mineralogy and the physical characteristics of the uppermost surface layer (e.g., Emery et al., 2006; Vernazza et al., 2012). The emissivity spectrum of an asteroid (see Fig. 5.8) is obtained by dividing the measured Spectral Energy Distribution (SED) in the thermal infrared by the modeled thermal continuum and multiplying this ratio by the assumed bolometric emissivity. Hence, to obtain a good emissivity spectrum a very high signal-to-noise ratio (SNR) SED is needed. In addition, ground-based MIR observations are limited by strong telluric absorptions and high levels of rapidly varying background emission. Because of these factors, the history of asteroid emissivity spectroscopy is relatively recent and the number of observed objects is relatively small. The recent development of high-sensitivity thermal 2D detectors and of space observatories have brought significant improvements and high SNR emissivity spectra have been obtained of tenths of asteroids.

The first emissivity spectra of asteroids were reported in a ground-based study by Feierberg et al. (1983). The Infrared Space Observatory (ISO) detected emissivity features in several asteroids (Barucci et al., 2002; Dotto et al., 2002, 2004; Müller and Blommaert, 2004).

The largest number of asteroid emissivity spectra with the best SNR in the 5.2 - 38 μm range was obtained with the *Spitzer* Space Telescope using the Infrared Spectrograph (IRS). From the ground, Lim et al. (2005) observed 29 asteroids in the 8 - 14 μm spectral region, identifying spectral structure in two of these: (1) Ceres and (4) Vesta. A large fraction of the mid-infrared (MIR) spectra of primitive asteroids show emissivity structures similar to those clearly detected in the spectra of three Trojan asteroids observed by NASA's *Spitzer* Space Telescope (Emery et al. 2006), shown in Fig. 5.8. Such emissivity structures have also been observed in at least 24 primitive asteroids including (10) Hygiea (Barucci et al., 2002), (308) Polyxo (Dotto et al., 2004), (21) Lutetia (Barucci et al., 2008), (65) Cybele (Licandro et al. 2011), plus 20 Themis and Veritas asteroids observed (Licandro et al. 2012, Landsman et al. 2016).

Most of the emissivity features in the MIR region are produced by silicate minerals. This region contains the Si–O stretch and bend fundamental molecular vibration bands (around 9–12 and 14–25 μm, respectively). Interplay between surface and volume scattering around these bands creates complex patterns of emissivity highs and lows which are very sensitive to, and therefore diagnostic of, mineralogy as well as grain size and texture. In particular, the shape and contrast of a broad spectral feature near 10 μm is sensitive to grain size, porosity and mineralogy, and also provide a link between asteroids, comets, Interplanetary Dust Particles, (IDPs) and meteorites (e.g., Emery et al., 2006; Vernazza et al., 2012; Vernazza et al. 2015). Interestingly, the emissivity spectra in primitive asteroids resemble emission from cometary comae (Emery et al. 2006) and cometary nuclei (Kelley et al., 2017) more closely than any of the powdered meteorites, minerals, or their mixtures, and regolith analogs; these links with the composition of comets are consistent with a proposed continuum between primitive asteroids and comets (Gounelle 2012). Emery et al. (2006) suggest that the surface layer producing these features is either a very under-dense, fairy-castle-like structure, or fine-grained silicates imbedded in a matrix of material relatively transparent in the mid-IR. Chondritic, porous IDPs can contain as much as 90% void space; perhaps on a low-gravity surface such a structure is likely. The similarities with the spectra of comet nuclei reported by Kelley et al. (2017) also supports the hypothesis proposed by Licandro et al. (2012) that the presence of a dust mantle on the surface of several of the observed Themis family asteroids is consistent with past water ice sublimation. The hypothesis of a very under-dense structure (Emery et al. 2006) is supported by the laboratory results described in Vernazza et al. (2012); they show that it is possible to reproduce the spectra of primitive asteroids with emissivity features by preparing the samples suspending meteorite and/or mineral powder (<30 μm) in IR-transparent potassium bromide (KBr) powder. This implies a high porosity (>90%) for the first millimeters of the asteroid surface. The large surface porosity inferred from the mid-IR spectra of certain asteroids is also implied by the determination of their thermal inertia and radar albedo (Vernazza et al. 2012; Landsman et al. 2018). As illustrated by Vernazza et al. (2012), information on the composition of the asteroid's surface can be obtained using a spectral decomposition model (Figs. 5.9 and 5.10) previously used to study the composition of cometary comae and protoplanetary disks (Lisse et al. 2006; Olofsson et al. 2010; Morlok et al. 2010).

Emery et al. (2006) did not find significant differences between the spectra of the three observed Trojans; however, clear differences have been found among the spectra of primitive asteroids, in particular in the emission plateau at 9 -12 μm. Licandro et al. (2012) and Landsman et al. (2016) presented 5–14 μm spectra of a sample of Themis and Veritas family asteroids obtained with NASA's Spitzer Space Telescope. Landsman et al. (2016) found that all the observed Themis-family asteroids (11) and six of the nine observed Veritas-family asteroids have a statistically significant 10-μm silicate emission band, with a spectral contrast that ranges from 1–8%. This is lower than the 10% amplitude of the silicate emission observed in the Trojans. They conclude that the lower amplitude of the silicate emission in the case of Themis and Veritas family asteroids – compared with that of Trojan asteroids – can be attributed to larger dust particles, to a slightly denser structure, and/or to a lower silicate dust fraction.

Differences in the shape of the 9-12 μm emission plateau can be used to identify differences in the mineral composition of the regolith. For example, McAdam et al. (2015) provide a method to remotely determine the alteration state of asteroids using the MIR wavelength region. They found that the emissivity minimum around 12 μm observed in the spectrum of CM and CI carbonaceous chondrites is correlated with the amount of phyllosilicates contained in the meteorites; the feature occurs at shorter wavelengths (near 11.4 μm) for more aqueously altered meteorites and longer wavelengths (near 12.3 μm) for less altered meteorites. Carbonaceous chondrites are considered to be representative of at least some primitive asteroids. Carbonaceous chondrites experienced extensive interactions with water during the first 5 myr of the Solar System's evolution (de Leuw et al., 2010), which produced aqueous alteration while in their asteroid parent bodies (e.g. McSween, 1979, 1987; Bunch and Chang, 1980; Tomeoka and Buseck, 1985; Tomeoka et al., 1989). This alteration is triggered when radioactively decaying elements heat the asteroid, melting the co-accreted water ice. As the meteorites exhibit variations in degree of alteration, one can expect minerals with different degrees of aqueous alteration on the surfaces of primitive asteroids. McAdam et al. (2015) used the relation between the position of the 12-μm minimum and the volume of phyllosilicates in carbonaceous chondrites to analyze the emissivity spectrum of (24) Themis and concluded that its regolith composition is similar to that of less aqueously altered meteorites. Applying this method Landsman et al. (2016) concluded that the regolith on their sample of Themis and Veritas family asteroids is also similar to the less aqueously altered meteorites. Both of these results are consistent with observations of 24 Themis in the 2 to 4-μm region, where the structure of the 3-μm band was indicative of water ice instead of hydrated minerals (Campins et al. 2010, Rivkin and Emery 2010).

Further compositional differences between different classes of primitive asteroids can be observed when applying a spectral decomposition model to their emissivity spectra. Using this method, Vernazza et al. (2015) show that there are remarkable differences between BCG-, PD-type asteroids, and comets. They show the dust observed in these surfaces is well sampled in our collections of IDPs, and actually better than in the collection of spectra of carbonaceous chondrites, thus confirming earlier suggestions/predictions (e.g., Bradley et al. 1996; Dermott et al. 2002 and references therein). Vernazza et al. (2015) show that BCG-type asteroids spectra are compatible with those of pyroxene-rich Chondritic Porous IDPs. The spectra of P- and D-type asteroids can be fit with a mixture of pyroxene-rich and olivine-rich Chondritic Porous IDPs.

Comets, on the other hand, appear mostly compatible with olivine-rich Chondritic Porous IDPs and to a lesser extent with a mixture of pyroxene-rich and olivine-rich Chondritic Porous IDPs. The Vernazza et al. (2015) results imply that IDPs sample a larger fraction of main belt asteroids than meteorites. Reciprocally, they suggest that asteroids are the likely parent bodies of a large fraction of IDPs.

After *Spitzer* and current ground-based instrumentation, the next leap in telescopic capability in the mid-infrared is the James Webb Space Telescope (JWST). JWST's Mid-Infrared Instrument (MIRI) provides low-resolution spectroscopy ($\lambda/\Delta\lambda \sim 100$) from 5-14 µm and medium-resolution spectroscopy ($\lambda/\Delta\lambda \sim 2200\text{-}3500$) and imaging in nine bands from 5-28 µm. The sensitivity of MIRI will allow compositional studies of much smaller objects than currently available (Rivkin et al. 2016).

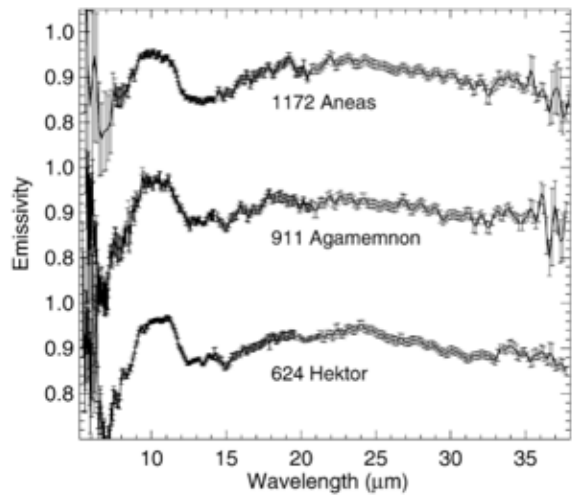

**Figure 5.8**. From Emery et al. (2006), the emissivity spectra of Trojan asteroids (424) Hektor, (911) Agamemnon, and (1172) Aneas created by dividing the measured SED by the best-fit Standard Thermal Model (STM) for each object.

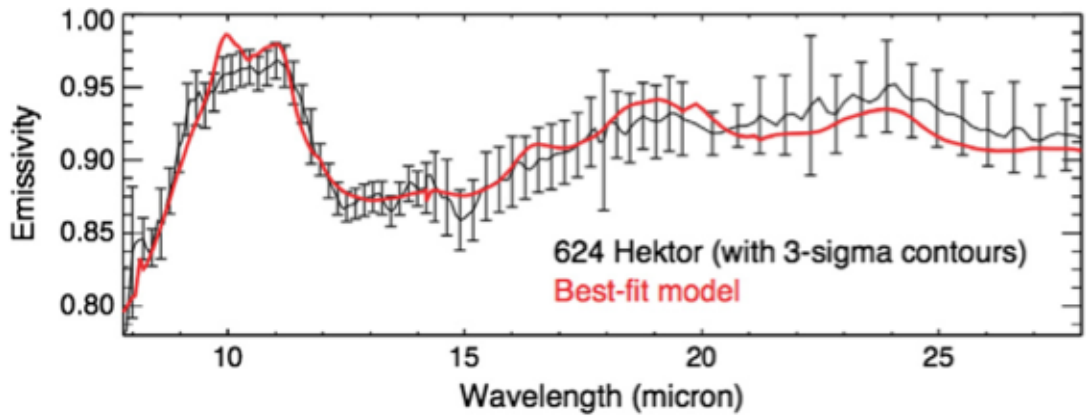

**Figure 5.9**. A comparison between the spectrum of asteroid Hektor's spectrum and the best-fit model obtained by Vernazza et al. (2012) using crystalline olivine, crystalline pyroxene, amorphous olivine, amorphous pyroxene and amorphous carbon with abundances in mass of 0.68, 0.05, 2.19, 0.07 and 97.01 %, respectively.

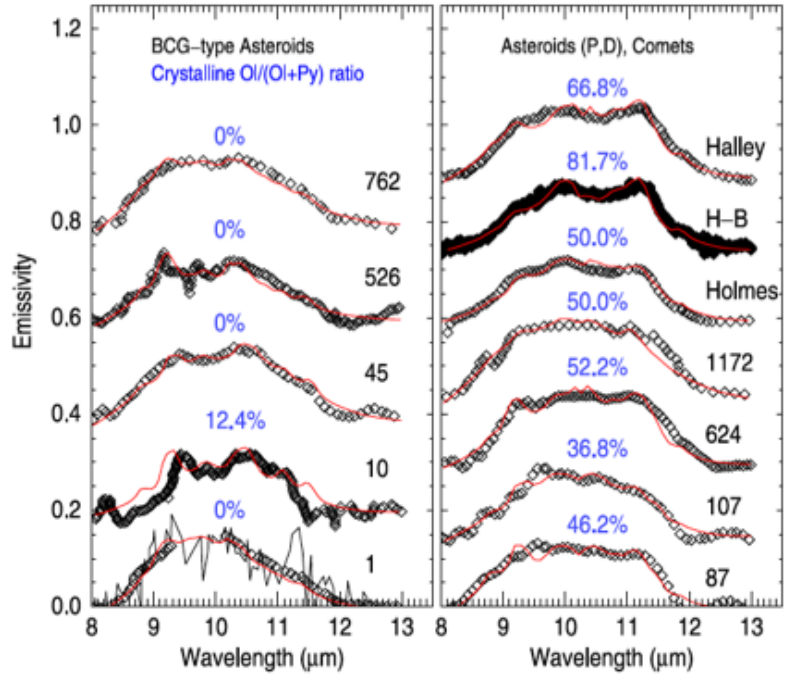

**Figure 5.10**. The composition of B, C, G, P and D-type asteroids, and of comets constrained using a Rayleigh scattering model from Vernazza et al. (2015). The best fits were obtained using four end members only (olivine, enstatite, olivine-like glass, pyroxene-like glass). The relative abundance of crystalline olivine to crystalline silicates (olivine+pyroxene) is directly indicated on the plot. (Left): BCG-type asteroids (Cohen et al. 1998; Barucci et al. 2002; Takahashi et al. 2011; Licandro et al. 2012; Marchis et al. 2012) have a crystalline ol/(ol+px) ratio in the 0–0.15 range. Most objects require no olivine. (Right): P-type asteroids (87) and (107) Marchis et al. (2012) have a crystalline ol/ol(px) ratio in the 0.3–0.5 range whereas the remaining objects (D-types and comets; Campins & Ryan 1989; Hanner et al. 1997; Emery et al. 2006; Harker et al. 2011) possess a crystalline ol/ol(opx) > 0.5.

**5.6 Effects of Space Weathering on Spectral Observations of Primitive Asteroids**

Space weathering (SpWe) is defined as the changes in the optical properties of the surfaces of airless bodies caused by solar wind and cosmic ray irradiation and by micrometeoroid bombardment (Brunetto et al. 2016). We start our discussion of SpWe with the effects at VNIR wavelengths because VNIR is the best-studied region. Until recently, laboratory experiments to simulate the effects of space weathering were focused on the study of silicates and ordinary chondrites. Dramatic darkening and reddening of the VNIR spectra were observed and attributed to space weathering, explaining the mismatch between ordinary chondrites and S-type asteroids slope distributions (e.g., Pieters et al. 2000; Marchi et al. 2005). While the effects of SpWe in the VNIR spectra of silicate-rich surfaces are quite clear (darkening and reddening), things appear more complicated when considering dark or primitive asteroids: weak reddening trends have been observed by Lazzarin et al. (2006) while Nesvorný et al. (2005) reported a bluing effect. The irradiation experiments of dark carbonaceous chondrites show different trends, with spectral modifications in the VNIR range correlated with the initial albedo/composition. More specifically, reddening was observed after ion and/or laser irradiation of CV3, CO3, and CM chondritic meteorites (Brunetto et al. 2014, Lantz et al. 2013, 2015), while a flattening or blueing effect was observed in the case of the Tagish Lake meteorite, a C2 ungrouped chondrite with affinities with the CM and CI chondrite groups (Hiroi et al. 2013). Lantz et al. (2017) proposed a model for space weathering effects on low albedo objects, where those with initial albedo between 5 and 9% do not show SpWe effects in the visible range; those with lower albedo show blueing, while those with higher albedo show reddening. Lantz et al. (2017) also report a SpWe shift toward longer wavelength in the bands due to phyllosilicates (near 3 and 10 μm) and other silicates (near 10 μm). The variations in the spectra of carbonaceous chondrites during SpWe laboratory experiments are always smaller than those observed on silicates or ordinary chondrites, and are comparable to variations due to sample preparation and other effects (e.g. grain size). All this may explain why SpWe trends for dark asteroids are less pronounced or absent.

Although there is less information about SpWe on primitive surfaces in the NUV than in the VNIR range, the effect of SpWe on the spectra of primitive asteroids and carbonaceous chondrites appears more pronounced. Hendrix and Vilas (2006) found that space weathering of S-class asteroids and the Moon manifests itself at NUV wavelengths as a bluing of the spectrum and its effects appear with less weathering, i.e., the NUV region seems to be a more sensitive indicator of weathering and thus, age. They compared spectroscopic data (0.22-0.35 μm) from International Ultraviolet Explorer of S-type asteroids with laboratory measurements of meteorites to obtain their results. The UV spectral bluing is most prominent in the 0.3-0.4 μm range, which was confirmed by Kanuchova et al. (2015). They analyzed the NUV (0.2-0.4 μm) laboratory reflectance spectra of ion-irradiated silicates and meteorites as a simulation of solar wind ion irradiation, finding a strong bluing effect at NUV wavelengths. They attributed this bluing to the formation of iron nanoparticles accompanied by amorphization of surface silicates. According to these authors, "*it will be important to include in future studies the NUV range both in the observations of asteroids and in the laboratory spectra of meteorites*".

Finally, studying SpWe effects on primitive asteroids in collisional families could be used to estimate the rate of SpWe. A collisional family is a group of fragments ("members") generated by a disruptive impact event on the parent body. This event may expose new material that, although originally unweathered (not altered by exposure to the space), will be altered by SpWe over time. Initial studies of primitive asteroid families with different ages, such as the Themis, Beagle, and Veritas families, show spectral differences. However, the cause for these differences between families is not clear; SpWe may not be necessary to explain these spectral differences, but it cannot be ruled out (Nesvorný et al. 2005; Ziffer et al. 2011; Campins et al. 2012; Fornasier et al. 2016).

In the inner-belt, there are interesting new results from visible spectra of the Clarissa family (Pinilla-Alonso 2017; Morate et al. 2018). Figure 5.11 (adapted from Pinilla-Alonso et al. 2017 and Morate et al. 2018) compares the spectral types in the Clarissa and Polana families. The spectral types present in both families are similar (B, C and X), and neither family shows the 0.7-µm feature. Based on these two characteristics, one can say that Clarissa is Polana-like and not Erigone-like (the two spectrally distinct groups of primitive asteroids identified so far in the inner belt, see Section 5.2.2). However, the Clarissa family is considerable younger, at less than 100 million years, than the Polana family at approximately 2000 million years (Nesvorný et al. 2015), and there are subtle yet significant spectral differences between these two families. These differences are consistent with the space weathering trend suggested by Lantz et al. (2015 and 2017), where the lowest-albedo primitive material gets bluer with space exposure age. In fact, the younger Clarissa family has a lower fraction of the bluer "B" spectral class and a higher fraction of the red "X" spectral classes than the older Polana family (Figure 5.11). This tantalizing agreement between observations and laboratory simulations is preliminary; however, it has testable implications for Bennu and Ryugu, where older terrains would be expected to be bluer than younger surfaces. More general predictions of the expected spectral characteristics of Bennu and Ryugu, based on the spectral diversity observed in inner-belt primitive asteroids are discussed in de Leon et al. (2018).

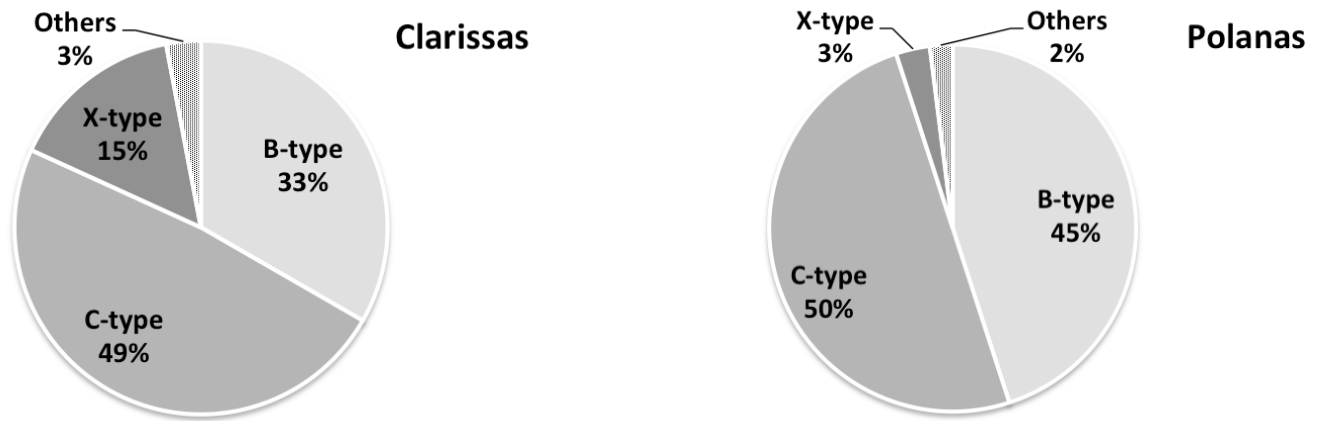

**Figure 5.11**. Taxonomical groups in the Clarissa and Polana families, adapted from Pinilla-Alonso et al. (2017) and Morate et al. (2018). The largest group in both families is C-type, with approximately 50% of the observed asteroids; however, the younger Clarissa family has a lower fraction of the bluer "B" spectral class and a higher fraction of the red "X" spectral classes than the older Polana family. These differences are consistent with the space weathering trend suggested by Lantz et al. (2015 and 2017), where the lowest-albedo primitive material gets bluer

with space exposure age. This tantalizing agreement between observations and laboratory simulations is preliminary; however, it has testable implications for Bennu and Ryugu.

In order to test the albedo predictions of Lantz et al. (2015 and 2017), we also looked at the distribution of visible geometric albedo of the Polana and Clarissa families; however, we could not detect any statistically significant differences. The distributions appear slightly different, and the Polanas have an average albedo of $0.046^{+0.015}_{-0.010}$ (1-σ), whereas the Clarissas have a slightly lower average of $0.040^{+0.02}_{-0.01}$. However, the sample of Clarissas is too small (only 19 objects, versus more than 600 Polanas) for this difference to be statistically significant, so more albedos would be needed to confirm any difference between these two populations. We obtained the albedos from WISE /NEOWISE data and tabulated absolute asteroid magnitudes (see Alí-Lagoa & Delbo' 2017 for details).

### 5.7 Summary

In this work we reviewed the observed spectral characteristics of primitive asteroids; below, we summarize the main results.

- There is clear spectral diversity among primitive asteroids. Diagnostic features from the UV to the mid-infrared are indicative of composition, particle size and other surface characteristics.

- The UV (~0.1-0.4 μm) can be diagnostic of carbon-rich compounds and hydrated silicates in primitive asteroids. In particular, a feature centered at ~0.2 μm has been associated with graphite grains or hydrogenated amorphous carbon, while a feature centered at ~0.43 μm is associated with ferric iron resulting from aqueous alteration.

- The most prominent feature in the visible spectra (~0.5-0.9 μm) of primitive asteroids is the 0.7-μm absorption. This feature is a valuable proxy for hydration; whenever the 0.7-μm band is present, the 3-μm band is also seen; the inverse is true in only half of the cases. Hence, a lack of a 0.7-μm feature does not necessarily mean that a surface is anhydrous.

- In the near-infrared region (~0.8-2.5 μm), the spectra of primitive asteroids are relatively featureless and exhibit a broad range of slopes, from negative (blue) to positive (red) slopes.

- The 3-μm absorption is most diagnostic of water ice, water-bearing materials, and surficial OH. Studies performed on carbonaceous chondrites show that a strong absorption edge near 2.7 μm is indicative of phyllosilicate composition and aqueous alteration.

- Most of the emissivity features of primitive asteroids in the mid-infrared region (5 – 40 μm) are produced by silicate minerals. The emissivity minimum around 12 μm observed in the spectra of some carbonaceous chondrites is correlated with the amount of

- phyllosilicates contained in the meteorites. This feature shifts from ~ 11.4 μm to 12.3 μm as the degree of aqueous alteration decreases.
- New mid-infrared connections between primitive asteroids and interplanetary dust particles indicate that the latter sample a larger fraction of main belt asteroids than meteorites and are consistent with a proposed continuum between primitive asteroids and comets.
- The effects of space weathering on the spectra of primitive asteroids are more complex than what is seen for ordinary chondrites and S-type asteroids. Irradiation experiments of dark carbonaceous chondrites show different trends, with spectral modifications in the visible and near-infrared range correlated with the albedo/composition of the sample.
- Primitive asteroids are present throughout the asteroid belt, including the inner-belt, which is the main source of NEAs. Hence, we can expect diversity among primitive NEAs, including spacecraft-accessible ones. This diversity may even be apparent within individual primitive NEAs. Soon we will have the opportunity to explore this possibility with *in situ* and sample-return studies of at least two objects: (101955) Bennu and (162173) Ryugu.
- So far, Inner-Belt primitive families fall into two compositional groups: Erigone-like (hydrated and spectrally diverse) and Polana-like (no 0.7-μm absorption and spectrally homogeneous)
- Both sample-return targets, Bennu and Ryugu, most likely originated in the Polana family. Like the Polana family, these NEAs do not show a 0.7-μm feature (at least not in their ground-based disk-integrated spectra available to date), they are spectrally compatible with the Polana spectral types (B and C-types) and are dynamically consistent with a Polana origin.
- An agreement between observations of inner-belt families and laboratory simulations of space weathering has testable implications for Bennu and Ryugu: older terrains would be expected to be bluer than younger surfaces

**Acknowledgements**

H.C. acknowledges support from NASA's Near-Earth Object Observations program and from the Center for Lunar and Asteroid Surface Science funded by NASA's Solar System Exploration Research Virtual Institute at the University of Central Florida. J.dL. acknowledges support from the Spanish 'Ministerio de Economía y Competitividad' under the 2015 Severo Ochoa Program SEV-2015-0548. J. L. and J.dL. acknowledge support from the AYA2015-67772-R (MINECO, Spain). V.A-L.: acknowledges that the research leading to these results has received funding from the European Union's Horizon 2020 Research and Innovation Programme, under Grant Agreement no. 687378. We thank N. Pinilla-Alonso and H. Frechette for valuable input.